\documentclass[%
 reprint,
 amsmath,amssymb,
 aps,r]{revtex4-1}

\usepackage{graphicx}
\usepackage{dcolumn}
\usepackage{bm}
\usepackage{epsfig}
\usepackage{hyperref}
\usepackage{float}

\usepackage{color}

\begin{document}

\preprint{APS/123-QED}

\title{Pressure anisotropy and small  spatial scales  induced by velocity shear}
\author{D. Del Sarto}
 \affiliation{  Institut Jean
Lamour, UMR 7198 CNRS - Universit\'e de Lorraine, France}
\email{daniele.del-sarto@univ-lorraine.fr}
\author{F. Pegoraro}%
\author{F. Califano}%
\affiliation{Physics Department and CNISM, Pisa University,
Italy}%

\begin{abstract}
  Non-Maxwellian metaequilibria can  exist in low-collisionality  plasmas as evidenced by  satellite and laboratory measurements. By including  the full pressure tensor dynamics  in a fluid plasma model, we show that a sheared velocity field  can provide an effective  mechanism that makes an initial isotropic  state anisotropic and  agyrotropic. We discuss how the propagation of  magneto-elastic waves  can affect  the pressure tensor anisotropization and its spatial  filamentation which are due to the action of both the magnetic field and flow strain tensor.
 We support  this analysis by  a numerical integration of the nonlinear equations describing the pressure tensor evolution.

\end{abstract}

\pacs{Valid PACS appear here}
\maketitle


The aim of this Letter is to show that a sheared velocity field in a weakly collisional, magnetized plasma 
drives a macroscopic pressure anisotropization in the plane of the velocity strain tensor.  This represents a general mechanism when  collisional relaxation is absent or slow that causes   part of the kinetic energy  of the  plasma flow to be   locally transformed into anisotropic ``internal energy''.   This energy conversion implies  that  shear flows do not  affect the plasma dynamics only through the fluid destabilization  of  Kelvin-Helmholtz (KH) modes \cite{KH} or  by breaking the correlation length of unstable modes \cite{Palermo,Transp_barrier}  responsible e.g.,  for anomalous  energy transport in magnetically confined plasmas, but  can lead to the onset of additional phase space instabilities driven by the induced pressure anisotropy.

In   magnetized  plasmas  the fast  particle gyromotion in  a sufficiently strong  field   makes the pressure tensor  isotropic  in the plane  perpendicular to the magnetic field  direction but allows for different parallel and perpendicular pressures (gyrotropic pressure as is the case for the double-adiabatic  or CGL\cite{CGL}  closure). On the contrary, the fluid strain $\Sigma_{ij}\equiv\partial u_i/\partial x_j$   in the  sheared fluid  velocity ${u_i({\bm x})}$  has a twofold effect:  first, through its rotational component it   combines or competes  with the gyrotropic effect due to the magnetic field, second it   induces pressure anisotropy   (agyrotropic pressure) in the plane perpendicular to the magnetic field (taken  to coincide with the velocity shear plane)  through its incompressible rate of shear (its symmetrical traceless component).

Here we discuss   the role of the  flow strain in  the  dynamic equations of the  full pressure tensor as obtained from the second  moment of Vlasov Equation (VE), thus going beyond both the   CGL closure and the Finite Larmor Radius (FLR) corrections\cite{Cerri_1}  approach. We  focus in particular  on the dynamics of the full pressure tensor within a 1-fluid description of a dissipationless magnetized plasma and show how the propagation of  ``{\it magneto-elastic}''
 waves  can affect  the pressure anisotropization and  { small spatial scale formation}  due to the interplay between the gyrotropic and  the non-gyrotropic dynamics induced by the magnetic field and by the strain  tensor.

Non-Maxwellian states, sometimes exhibiting pressure agyrotropy\cite{Astudillo,Posner,He}, are observed both experimentally\cite{Marsch,CLUSTER,Astudillo,Posner,He,Tu,Wind,Aunai1,LEIA} and in Vlasov simulations\cite{Servidio1,Galeotti}; also the role of a shear  flow in affecting the kinetic properties    of a collisionless or weakly collisional plasma
  is well established experimentally.
 Even if  mechanisms based on the  CGL  paradigm are now almost acquired when explaining the main features of solar wind anisotropies\cite{Wind}, the correlation between the presence of a velocity shear and the extent of anisotropy in particle distributions is evoked both for the core protons in the fast solar wind\cite{Tu}  and in ``space simulation'' laboratory
 experiments\cite{LEIA}. A sheared convection velocity in the Earth  ionosphere is thought to play a role in the heating of ions and in the consequent plasma bulk upflow in the auroral region\cite{Aurorae}. Moreover, the presence of a velocity shear is known to play an important role in the enhancement of a variety of pressure  anisotropy-related plasma instabilities. The presence of a velocity shear in the near-Earth plasma sheet profile prior to a substorm expansion lowers the instability threshold of ion-Weibel modes in the geomagnetic tail\cite{Yoon}.  {The role of a vorticity-related velocity shear in generating a gyrotropic temperature anisotropy in ion temperature gradient (ITG) driven turbulence, was discussed in Ref.\cite{Transp_barrier} in connection with  the relaxation of transport barriers in tokamak plasmas. Finally, an anisotropic pressure tensor was shown to form during the nonlinear stage of the current-filamentation instability (CFI) arising in the presence of two opposite cold electron beams. This anisotropy was  shown to increase the threshold and growth rate of the reconnection instability developing on the shoulder of the CFI-generated magnetic structures\cite{Califano1}. These also develop in the presence of  radially inhomogeneous beams such as in high intensity laser-plasma interactions\cite{Califano2} and are measured in {laboratory} experiments\cite{inertial}.   \\
 In Ref.\cite{Kahn} sufficient conditions for the instability of electromagnetic waves in an unmagnetized Vlasov plasma with a sheared velocity distribution were studied, showing that an anisotropic pressure tensor leads to the instability of transverse perturbations.  This was applied in \cite{Albright} to study the steady state achievable when the shear-induced temperature anisotropy  balances the electron diffusion in the velocity space  due to the growth of static magnetic perturbations generated by the  pressure  anisotropy itself.  Anisotropic turbulence induced by a KH unstable velocity shear\cite{Werne} and by a Von Karman flow\cite{Marie} was also pointed out.\\
\\
We start from  the two-fluid equations of a collisionless magnetized  plasma, obtained by evaluating the moments of VE. We neglect  the  electron dynamics (i.e. $m_e/m_i\rightarrow 0$) and temperature, whereas the components of the full pressure tensor contribute to ion dynamics.  The $2^{nd}$ anisotropic ion  moment,  $\Pi_{ij}$, is closed by
\begin{equation}\label{eq:1}
\frac{\partial}{\partial t} \Pi_{ij}
\,=\,-\underbrace{ \mathcal{L}_{{\bf u}}(\Pi_{ij})
}_{|\nabla{\bf u}|\equiv\tau_{H}^{-1}}
\,+\,\underbrace{\mathcal{M}_{{\bf u}}(\Pi_{ij})}_{\Omega_{c}\equiv\tau_{B}^{-1}}
\end{equation}
where  $\Omega_{c}\equiv q|{\bm B}|/(m c)$   
 and we have introduced the linear 
operators $\mathcal{L}_{{\bf u}}(\Pi_{ij})\equiv {\partial_{k}}(u_k\Pi_{ij})+\Pi_{kj}{\partial_{k}}{ u_{i}}+\Pi_{ik}\partial_{k}{ u_{j}}$ and $\mathcal{M}_{{\bf u}}(\Pi_{ij})\equiv 
q/m
 \left(\varepsilon_{ilm}\Pi_{lj}B_{m}+\varepsilon_{jlm}\Pi_{il}B_{m}\right)$, and the corresponding  characteristic hydrodynamic ($\tau_{_H}\equiv|\nabla{\bf u}|^{-1}$) and magnetic ($\tau_{_B}\equiv \Omega_{c}^{-1}$) time scales.  {The simplifying assumption, frequently used in the literature \cite{Hesse,Khanna}, of neglecting in Eq.(\ref{eq:1}) the divergence  of the ion  
heat flux tensor, $Q_{ijk}\equiv \langle mn(v_i-u_i)(v_j-u_j)(v_k-u_k)\rangle$,}  is consistent with the geometrical configuration considered later in this letter at least until very short spatial scales in the plane perpendicular to the magnetic field are formed during the 
 nonlinear evolution. Typical closures of ${\bm \Pi}$ have been performed by means of a power expansion each in terms of some small parameter: including corrections in Eq.(\ref{eq:1}) due to a small collision time with respect to both $\tau_{_H}$ and $\tau_{_B}$, leads to Braginskii's gyroviscous FLR model\cite{Braginskii}, while a small $\tau_{_B}/\tau_{_H}$ leads to FLR gyrotropic corrections to  CGL equations\cite{Cerri_1,Khanna,FLR_CGL}. More recent FLR-Landau-fluid models\cite{FLR_LF} also retain Landau-fluid effects\cite{LandauFluid}.  
Here we consider a dissipationless  regime  and do not assume the ratio  $\tau_{_B}/\tau_{_H}$  to be small.  \\  Defining the matrices
${\bf B}_{ij}\equiv\varepsilon_{ijm}B_m$,  and  ${\bf W}_{ij}\equiv $ $ (\partial_i u_j-\partial_j u_i)/2$ 
that describe the rotation induced by the magnetic field and by the shear flow respectively, the strain traceless matrix ${\bf D}_{ij}\equiv(\partial_j u_i+\partial_i u_j)/2 - C {\bm\delta}_{ij}$, the compression $C\equiv(\partial_ku_k)/3$ and the derivative $d/dt\equiv \partial_t+u_k\partial_k$, Eq.(\ref{eq:1}) can be conveniently written  as
\begin{equation}\label{eq:2}
\frac{d}{dt}{\bm\Pi}
={[\mathcal{\bf B}+
{\bf W}, {\bm\Pi}]}-\{{\bf D},{\bm\Pi}\}-5C{\bm \Pi}
\end{equation}
where $[\,,\,]$ denotes commutator and $\{\,,\,\}$ anticommutator. The first r.h.s. term shows that the magnetic field ${\bm B}$ and the flow vorticity ${\bm\omega}$ {($\omega_i\equiv\varepsilon_{ijk}W_{jk}$)} combine to make ${\bm\Pi}$ to rotate around the axis of ${{\bf B}+{\bf W}}$. The perpendicular components rotate at twice the cyclotron frequency in absence of vorticity, or at twice the  fluid rotation frequency  in the unmagnetized case. If the axes of  ${\bf B}$ and  ${\bf W}$ are aligned the two frequencies add up if {${\bm B}\cdot{\bm\omega}>0$} and subtract if {${\bm B}\cdot{\bm\omega}<0$}. The last r.h.s. term of Eqs.(\ref{eq:2}) acts isotropically on ${\bm\Pi}$ while the second term can induce pressure anisotropization  whenever {${\bf D}$ is not zero.}

First, we consider a model system with an incompressible shear flow $u^0_y(x)$ constant in time (energy is thus constantly injected from outside) in the presence of a uniform and constant magnetic field along the $z$-axis. In this model the velocity strain and the vorticity have the same magnitude, ${\bf B}$ is uniform in space, the axes of ${\bf B}$ and ${\bf W}$ are aligned along $z$ while ${\bf D}$ has no $z$ components. Eq.(\ref{eq:2}) leads to a linear system in which the components of ${\bm \Pi}$ perpendicular to ${\bm B}$ are decoupled from those having a parallel component  {and their determinant  is constant}. Three eigenvalues are obtained  for the perpendicular components: $\gamma_0=0$, which corresponds  to a stationary mode    with 
$\Pi^{\gamma_0}_{yy}/\Pi^{\gamma_0}_{xx} =  \Omega'({x})/\Omega_{c} $ and  $\Pi^{\gamma_0}_{xy} =0$, 
and $\gamma_\pm=\pm 2i\sqrt{\Omega_{c}\Omega'(x)}$ {with}  $\Pi^{\gamma_\pm}_{yy}/\Pi^{\gamma_\pm}_{ xx} =  - \Omega'({x})/\Omega_{c} $  and $\Pi^{\gamma_\pm}_{xy} /\Pi_{xx}=  \pm  i\, \sqrt{\Omega'({x})/\Omega_{c}} $. Here   $\Omega'({x})\equiv\Omega_{c}+\partial_xu_y^0(x)$.   Provided $\Omega'({x}) >0$, the $\gamma_0$ mode can describe  an equilibrium solution of Eq.(\ref{eq:2}) (in agreement with the self-consistent equilibria discussed in \cite{Cerri_2})},  $\Pi_{yy}/\Pi_{xx} =  \Omega'({x})/\Omega_{c} $ and $\Pi_{xy} = 0$. The $\gamma_\pm $ 
modes  represent  either oscillations or  growing and damped modes depending on the sign of  $\Omega'({\bm x})$. { For  $\Omega'(x)>0$ the perpendicular pressure tensor components of an initial isotropic state with $\Pi_{xx}(x,0)=\Pi_{yy}(x,0)=P_{\perp}(x)$
oscillate in time   around a mean value
 given by  $\langle\Pi_{yy}(x,t)\rangle=(\Omega'(x)/\Omega_c)\langle\Pi_{xx}(x,t)\rangle=  (\Omega'(x)+\Omega_c) P_{\perp}(x)/(2\Omega_c)$ and $\langle\Pi_{xy}(x,t)\rangle=0$,  the amplitude of the oscillations  of $\Pi_{yy}(x,t)$ being $\partial_xu_y^0(x) P_{\perp}(x,0)/(2\Omega(x))$. } In Fig.\ref{Fig_eigenmodes} the profile of $\Pi_{yy}$ is shown at different times, for an initial pressure tensor $\Pi_{ij}=\delta_{ij}$, $B_z^0=1$ and  {$u_y^0=V_0 {\cos}(kx)$} with {$V_0=1.5$} and $k=1$. An important feature caused by the spatial inhomogeneity of the shear flow is the strongly inhomogeneous growth of the components of the pressure tensor,  as  regions where the evolution is oscillatory alternate, depending from the local sign of $\Omega'\Omega$, with regions of exponential growth occurring over a time scale $\tau_{_H}=(kV_0)^{-1}$. This gives rise to a spatially filamented pressure tensor. 

Second,   we consider the self-consistent (SC) case in which the flow  and the  e.m. fields evolve in time  according {to Eq.(\ref{eq:1}) and to the ideal MHD equations with an anisotropic ion pressure (note the  Hall  term  in Ohm's law to vanish  identically for  the examples discussed in this Letter). 
This system {conserves the total energy 
${E}^{tot}=\int d{\bm x}^3\left( \rho u^2/2 + B^2/8\pi  +\mbox{tr}\{{\bm\Pi}\}/2 \right)$,}   and depends on three dimensionless parameters
$\tau_H/\tau_B$ $ =(c_{_A}/c_{_H})(L_{_H}/d_i)$,   $(c_{_A}/c_{_H})^2$ and $(c_\perp/c_{_H})^2$ with $L_{_H}$ the scalelength of the configuration, $c_A$  the Alfv\`en velocity, $c_{_H} = L_{_H}/\tau_{_H} $ a measure of the flow velocity, $d_i\equiv c_{_A}/\Omega_{c}$ the ion skin depth, and $c_\perp\equiv P_{\perp}/\rho=c_s/\sqrt{2}$ with $c_s$ the   ``sound'' velocity  evaluated with respect to the initial ion pressure, assumed isotropic in the plane perpendicular to ${\bm B}$ \cite{comment_sound}. Two parameters only, $ (\tau_{_H}/\tau_{_B})/(c_{_A}/c_{_H})^2 $ and $(\tau_{_H}/\tau_{_B})/(c_\perp/c_{_H})^2$, rule the linear dynamics. The CGL-FLR limit of Ref.\cite{Cerri_1} is recovered in the low frequency limit $\partial_t\ll \Omega_c$ for $(c_{_H}/c_{_A})(d_i/L_{_H}) \ll 1$ and $d_i/L_{_H} \sim  (c_\perp/c_{_A})  d_i/L_{_H} \ll 1$, where no specific ordering for $c_\perp/c_{_A}$ is  assumed.

{In the SC case the anisotropization of the pressure tensor caused by the presence of an initially imposed  shear flow is limited by the  the reaction of the pressure tensor on the plasma flow which reduces its shear and by  the excitation of  nonlinear ``{\it magneto-elastic}''
 perturbations  that  tend to propagate  the  shear of the velocity flow outwards.}
The main features of these pressure tensor and velocity  perturbations can be understood by  referring to the linear waves described by  the  SC system that  propagate in  a uniform, homogeneous equilibrium with  ${\bm B}=B_0{\bm e}_z$, density $\rho_0$, {double adiabatic} pressure {$\Pi_{ij}=P_\perp\delta_{ij}+(P_{||}-P_\perp)B_iB_j/B^2$}, and   wave-vector ${\bf k}=k{\bm e}_x$. The resulting dispersion relation
\begin{equation}
\displaystyle{ (\omega^2-4\Omega^2 -k^2c_\perp^2)[\omega^2-k^2(c_{_A}^2+
 3c_\perp^2)]}
\displaystyle{ - 4\Omega^2k^2c_\perp^2=0},\\ \\
\label{DR} \end{equation}
 has a higher frequency branch HFB, $\omega^2_{h} \sim 4\Omega^2 {+2k^2c_\perp^2}$ and $\omega^2_h \sim {k^2(c_{_A}^2+3c_\perp^2)}$ for $k^2 \to 0$, and $k^2\to \infty$ respectively, and a lower frequency branch   LHB, $\omega^2_l \sim  {k^2(c_{_A}^2+2c_\perp^2)}$ and $\omega^2_l \sim   {k^2c_\perp^2}$ for $k^2 \to 0$, and $k^2\to \infty$.   The $k^2 \to 0$ limit of the LFB corresponds to the CGL form of  a perpendicular magnetosonic wave.  
In the limit of vanishing magnetic field the two branches are not dispersive and  reduce  to a longitudinal and to  a  transverse sound  mode propagating at phase velocities $\sqrt{3}c_\perp$ and $c_\perp$ respectively. 
{Note that  in this latter limit the SC system admits the propagation of finite-amplitude transverse waves where $u_y$ and $\Pi_{yy}$ satisfy the wave equation  $\partial_t^2 =c_\perp^2\partial_x^2$ and  the energy continuity equation $\partial_t( \rho_0 u_y^2 + \Pi_{yy})=-\partial_x(u_y\Pi_{yy})$.  
 When excited, these waves  may  carry  away and disperse an initially imposed shear flow $u_y(x)$  with spatial scale $L_{_H}$ on time scales of the order of $L_{_H} /c_{_A} \sim L_{_{H}}/c_\perp$, thus reducing the anisotropization  and the spatial inhomogeneity  of the pressure tensor forced by  the flow. This  propagation might  affect the onset   of the KH instabilities in a collisionless  weakly magnetized plasma. 
In the presence of  a strong magnetic field  these waves,  in particular the HFB,  couple  the $x$ and the $y$ components of the plasma velocity. In addition they become dispersive with the group velocity of the HFB going  to zero for $k\to 0$.

The nonlinear SC case has been integrated numerically starting from an isotropic initial condition with homogeneous density, ${\bm B}=B_0{\bm e}_z$ and ${\bm u}=u_y^0(x){\bf e}_y$, varying the  value of the ratios  of the three dimensionless parameters. In Figs.\ref{Fig_u_B_1}-\ref{Fig_Pi} we consider  the case with $u_y^0(x)=V_0\tanh(x/d_i)/\cosh^2(x/d_i)$ and $\tau_{_H}/\tau_{_B}= c_{_A}/c_{_H}=c_\perp/c_{_H}=1$.
The results obtained  can be qualitatively accounted for    by referring to  the  linear modes  described above where the initial shear velocity $u_y(x)$ is interpreted as an initial  perturbation. Note that, though its characteristic scale-length is  chosen of the order $d_i$, the initial Fourier spectrum  peaks around $kd_i\lesssim 1$ (Fig.\ref{Fig_u_B_1}). 
\\  The initial perturbation can be written as a superposition of the LFB and of the HFB. To  leading order in $kd_i\ll 1$  the polarization  vectors 
 components   in the $(u_x,u_y)$ basis    are $(1, -i ~ o(kd_i))$   and  $(1,-i)$   for  $\omega_l$ and  for $\omega_h$ respectively.
This implies that the  chosen initial perturbation corresponds to a superposition of the two branches with equal and opposite amplitudes and that  the time evolution of $u_y(x)$ 
is mainly determined by that of the HFB. 
This is consistent with the  results of the numerical integration and   explains why the  pressure agyrotropy  (Figs.\ref{Fig_Pi}), that  in our geometry is  mainly related  to the  spatial inhomogeneity of $u_y^0(x)$, tends to  remain at the original position  and not to be carried away at the  Alfv\`enic group velocity of the LFB, at least  until small spatial scales are formed which  are instead transported away efficiently by the HFB. For example, at $x=0$ Figs.(\ref{Fig_Pi}) show local peak agyrotropies $|\Pi_{yy}-\Pi_{xx}|/|\Pi_{yy}^0+\Pi_{xx}^0|$ of $0.12$, $0.22$ and $0.14$, and ``mean anisotropies'' $2\Pi_{zz}/(\Pi_{yy}+\Pi_{xx})$ of $1.02$, $2.35$ and $0.98$ for the  $\tau_{_A}/\tau_{_B}=0.1,1,10$ case respectively. On the contrary, both branches contribute to the evolution of $u_x(x)$  where the initial cancellation is removed as time evolves with the LFB component propagating outwards and the HFC  essentially mirroring, with an inverted  sign,  the behavior of $u_y$.\,  An increase of the  magnetic field $B_0$ has a double role:  on the  one  hand it tends to enforce
perpendicular gyrotropy 
while on the other, when the ratio $c_{_A}/c_\perp$ is increased,  the group velocity of the HFB decreases and the initial perturbation of $u_y$ remains longer  confined at its initial position. \\
The  fluctuations of $u_x$   being compressible   induce fluctuations of $B_z$ and $\Pi_{zz}$  consistent with  the magnetoacoustic polarization $\delta \Pi_{zz}/\Pi_{zz}^0=\delta B_z/B_z^0$ (non shown here).} 
\\
The interplay between the filamentation shown in Fig.\ref{Fig_eigenmodes} and the propagation of the  disturbances of the pressure tensor  result in the formation of fine-scale spatial structures.  The latter dominate for higher values of $B_0$ because of the  nonlinear steepening  of the  front of the propagating ``magneto-elastic waves'',  as evidenced by the {$u_x$ and} $\Pi_{xx}$ profiles in Figs.\ref{Fig_u_B_1}-\ref{Fig_Pi}.\\ The following points appear as particularly relevant in the analysis of the SC case: \\ 1) analogously to the CGL paradigm, the ratio between the average pressure in the perpendicular plane and the parallel component can vary  considerably (e.g. $\tau_{_H}/\tau_{_B}= 1$ case in Fig.\ref{Fig_Pi}),\\ 2)  agyrotropic anisotropization also occurs, though less pronounced as  the magnetic field increases,  since the flow energy may be preferentially transferred to one tensor component only \cite{comment_asymmetry} in the plane perpendicular  to the magnetic field direction.  \\ 3) The pressure anisotropization is spatially asymmetric, depending on the sign of ${\bm\omega}\cdot{\bm B}$, and tends to be localized in space while moving together with the initial velocity inhomogeneity which is carried away by the nonlinear magneto-elastic type wave-packets, \\ 4) a relaxation toward new non-MHD equilibria like those discussed  in \cite{Cerri_1} may be asymptotically achieved, even if, at the increase of $B_0$, the spatial inhomogeneity  ${\bm u}=u_y^0(x){\bm e}_y$  transferred to the relevant $\Pi_{ij}$ components takes longer to leave  the initially sheared region when it is peaked  at large wavelengths ($kd_i\lesssim 1$). 
\\The plasma dynamics  described  above explains why isotropic MHD equilibria  cease to be equilibria  in presence of a stationary sheared flow \cite{Cerri_1,Cerri_2} and why  an  initial anisotropic distribution function is needed to initialize kinetic simulations in presence of a velocity shear\cite{Belmont}. This can  affect  the onset  and development of the KH instability and may have implications for the evolution of transport barriers in tokamak turbulence\cite{Transp_barrier}. Another direct implication is for turbulence itself:  since small-scale spatial inhomogeneities are naturally developed during the direct cascade, we may expect  that isotropic turbulent states are not likely to exist whenever a full pressure tensor evolution is accounted for.  In particular, since non-negligible discrepancies with respect to the CGL closure become important when $\tau_{_{H}}^{-1}\sim\Omega_{c}$,  for $c_{_H}\sim c_{_A}$ (Alfv\`enic turbulence)  pressure anisotropies in the plane perpendicular to the magnetic field can be expected when velocity inhomogeneities are generated at a scale $L_{_H}\sim d_i$,
 apparently in agreement with the temperature anisotropization observed in Refs.\cite{Servidio1} in concomitance with the development of current and vorticity layers of thickness $\sim d_i$. \,
Finally we recall that  the occurrence of  an agyrotropic pressure tensor  is well documented in solar wind measurements\cite{Astudillo,He}, possibly correlated to  plasma flows,  see e.g. Ref.\cite{Posner}.

\begin{acknowledgments}
The authors acknowledge useful discussions with S.S. Cerri (IPP-Garching), T. Passot and P.L. Sulem (Obs. de la C\^ote d'Azur) and A. Tenerani (UCLA). Part of this work  was funded by the FR-FCM grants 1MHD.FR.12.05 and 1IPH.FR.13.22. 
\end{acknowledgments}

\begin{figure}
\centerline{\epsfig{figure=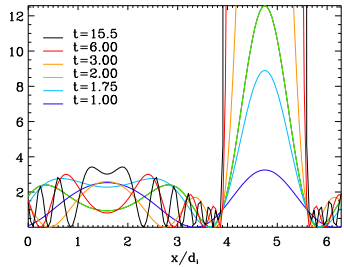,width=5.5cm}}
\caption{Evolution of $\Pi_{yy}(x,t)$   for  ${\bm B}=B_0{\bm e}_z$ and  constant ${\bm u}=(0, V_0\cos(x/d_i),0)$,  $\Omega_{c}\tau_{_H}=1$ and  $V_0=1.5 c_{_A}$. Both the exponential growth ($\Omega'(x)>0$) and the spatial filamentation of the oscillating solutions ($\Omega'(x)<0$) are visible. }\label{Fig_eigenmodes}
\end{figure}

\begin{figure}
\centerline{\epsfig{figure=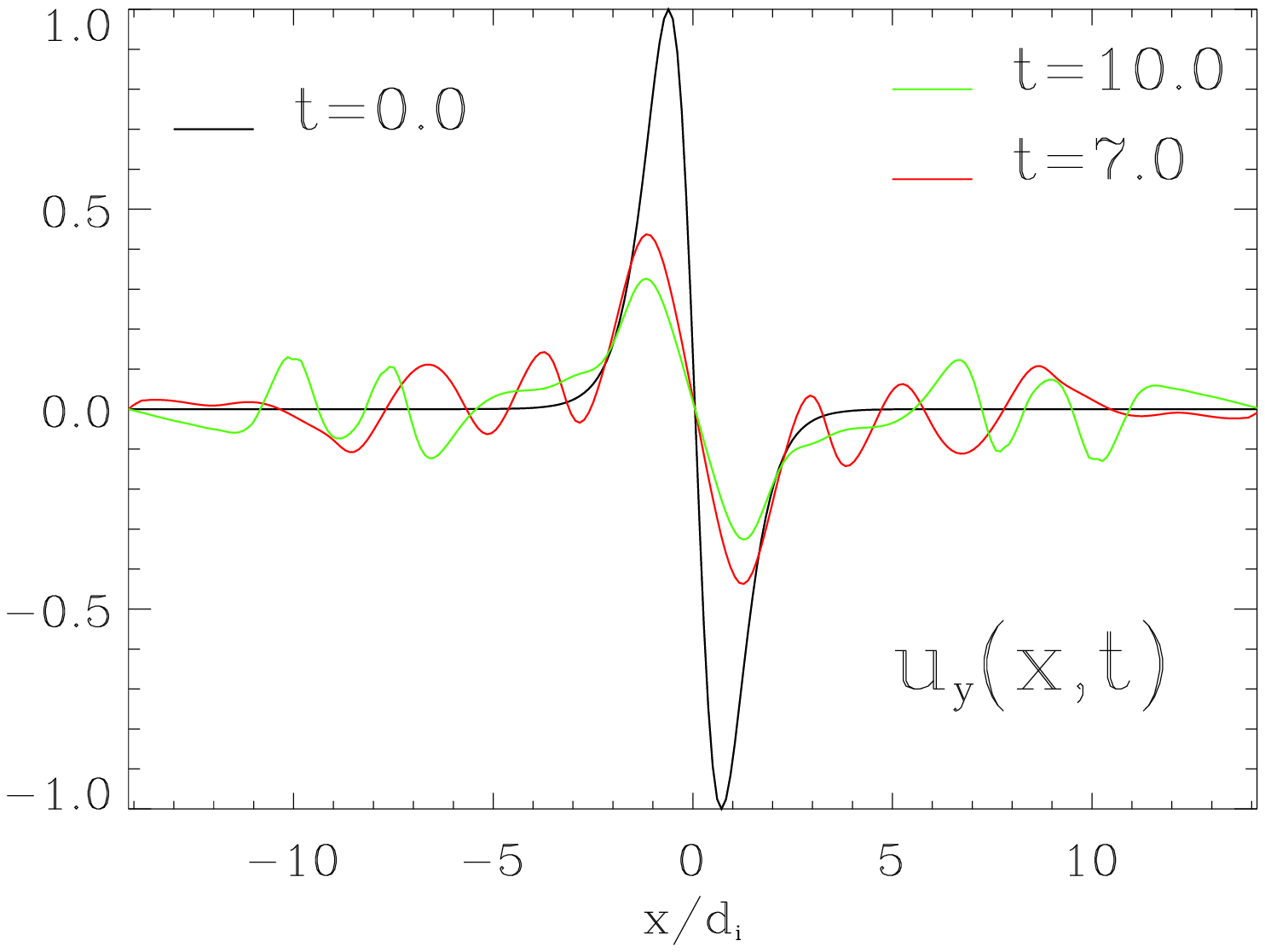,width=4.5cm},
\epsfig{figure=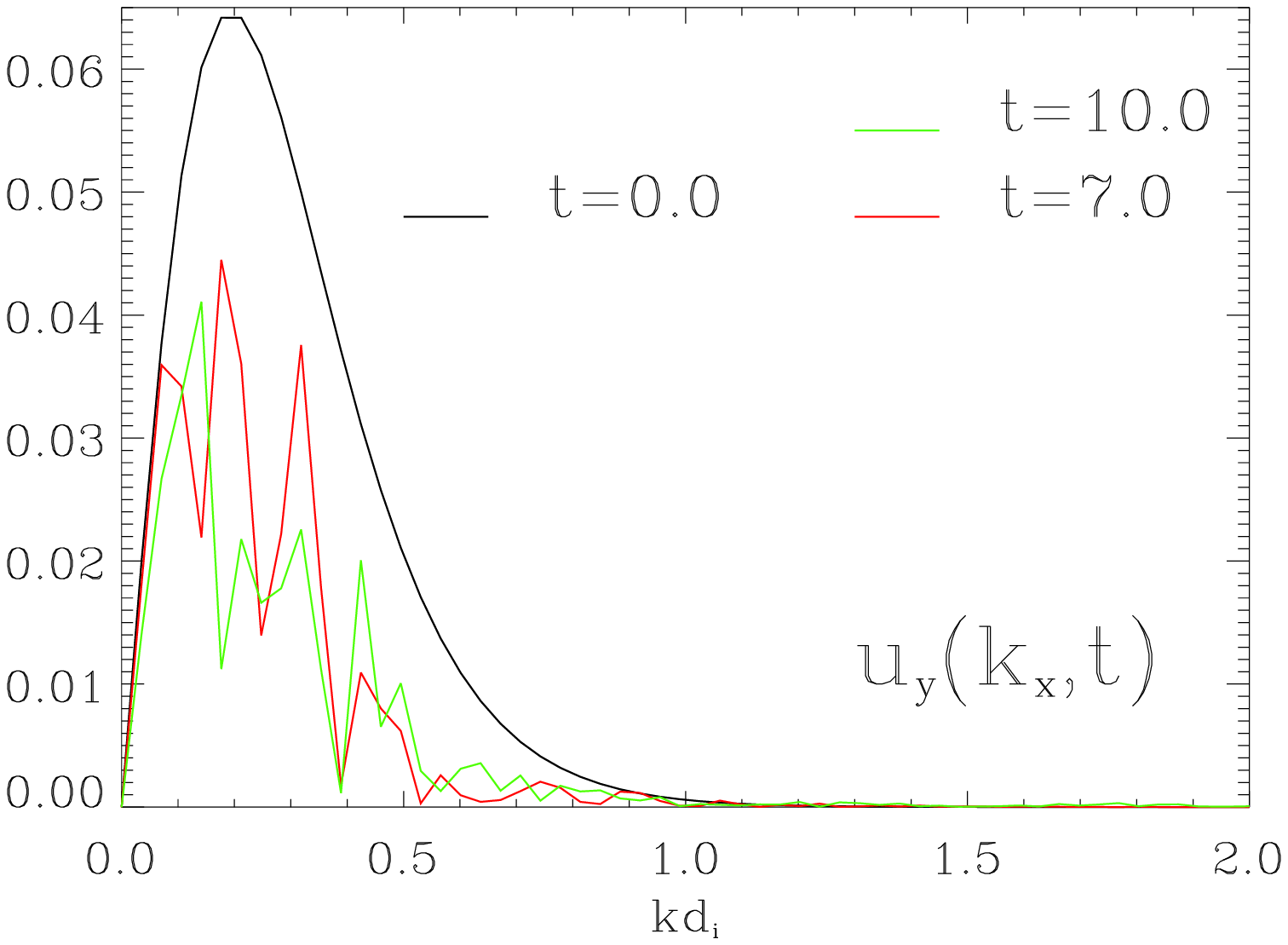,width=4.5cm}}
\centerline{\epsfig{figure=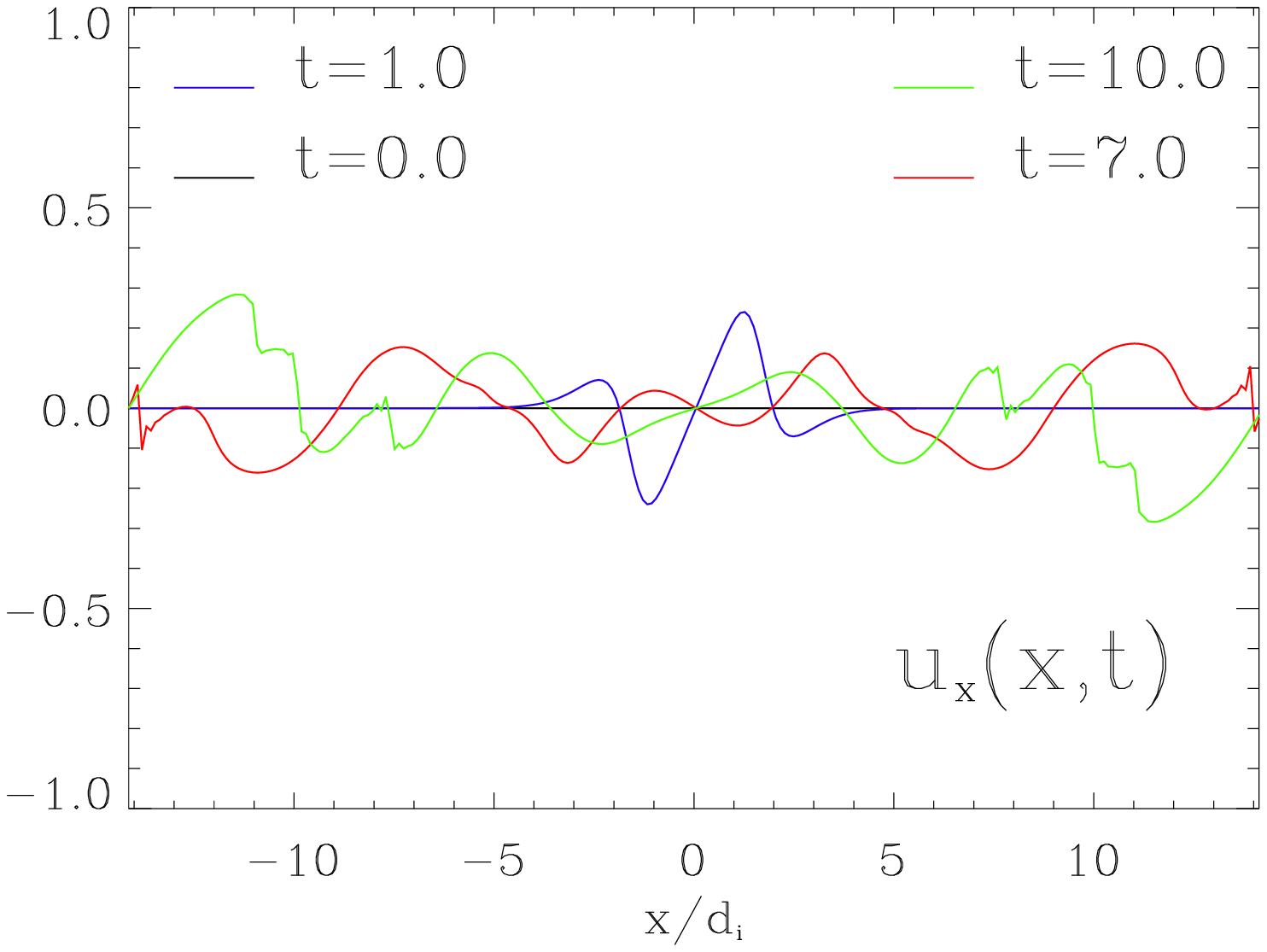,width=4.5cm},
\epsfig{figure=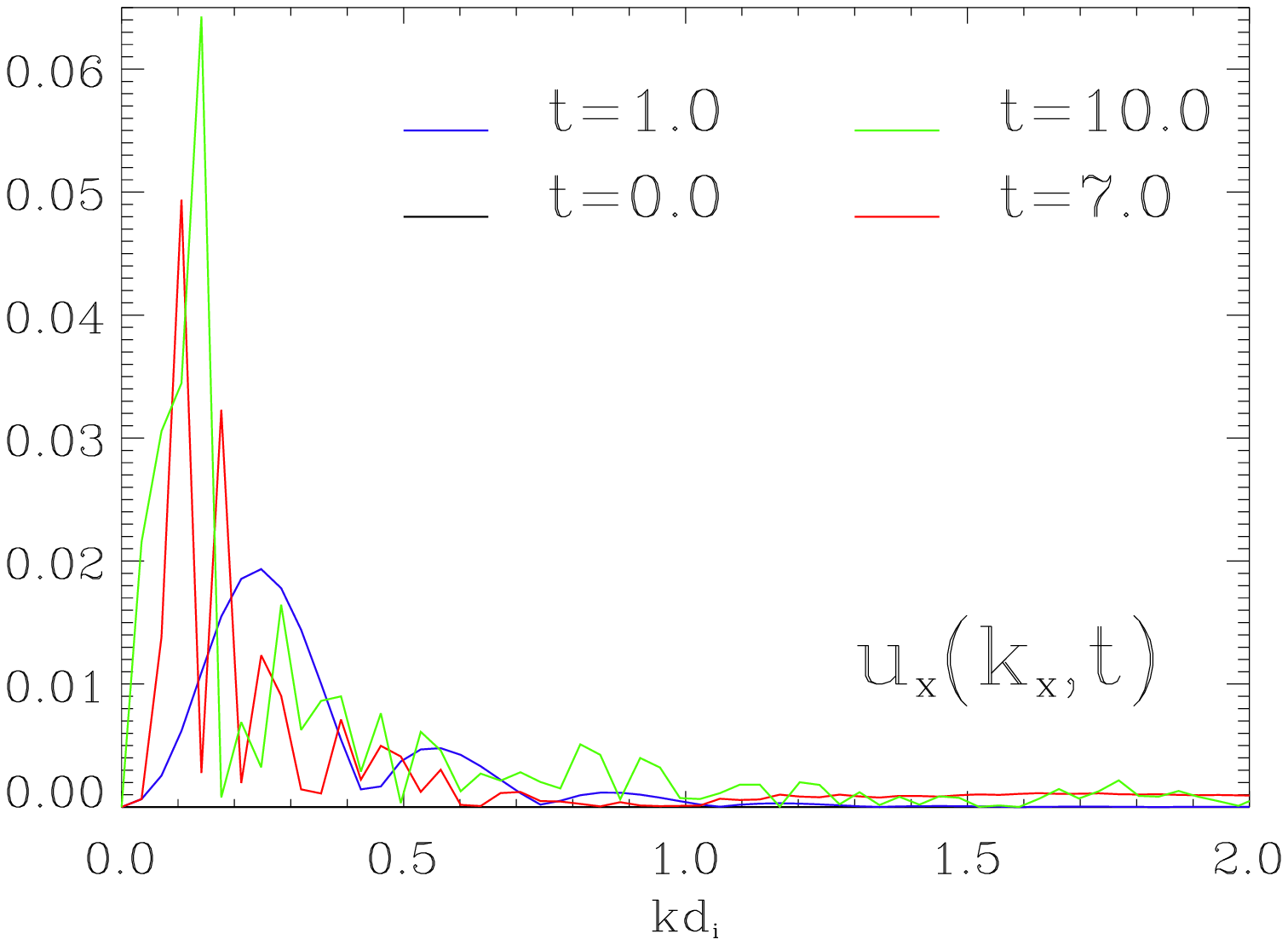,width=4.5cm}}
\caption{{Profiles of $u_x(x,t)$ and $u_y(x,t)$ (left) and their Fourier spectra  (right), for $c_{_H}=c_\perp=c_{_A}=1$; times in $\tau_{_H}=\tau_{_B}$ units.}}\label{Fig_u_B_1}
\end{figure}

\begin{figure}
\centerline{\epsfig{figure=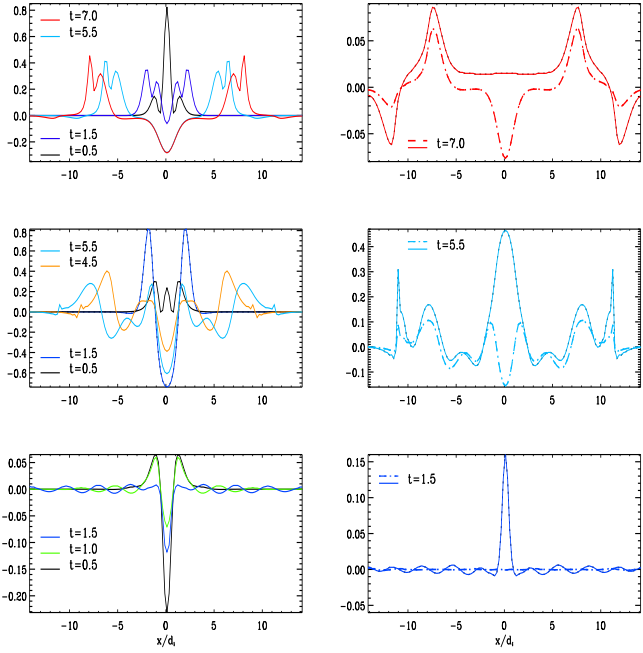,width=9.cm,height=9.cm}}
\caption{Spatial profiles of the local difference from the initial value for $\delta\Pi_{yy}(x,t) $ (left frames),    $\delta\Pi_{xx}(x,t)$ (solid lines, right frames) and $\delta\Pi_{zz}(x,t)$ (three dotted-dashed lines, right).  The initial pressures are uniform and isotropic ($\Pi_{ij}^0=\delta_{ij}$) with $c_\perp/c_{_H}=1$. As times are in $\tau_{_H}$ units, the nonlinear waves leave the box earlier in the bottom frames, each row  corresponding, from the top down, to $\tau_{_H}/\tau_{_B} = c_{_A}/c_{_H}= 0.1, 1$ and $10$ respectively (here $L_{_H}=d_i$).}\label{Fig_Pi}
\end{figure}


\begin{thebibliography}{99}
\bibitem{comment_sound} This is coherent with the definition of the sound velocity in the cold electron limit, where $c_s^2=2c_\perp^2=2P_{\perp}/\rho$, compatibly with the fact that in our 2D geometry the CGL evolution of $P_\perp$ is a polytropic with index $\Gamma=2$.
\end{thebibliography}
\end{document}